\pdfoutput=1

\documentclass[11pt]{article}

\usepackage[final]{acl}

\usepackage{times}
\usepackage{latexsym}

\usepackage[T1]{fontenc}

\usepackage[utf8]{inputenc}

\usepackage{microtype}

\usepackage{inconsolata}

\usepackage{soul}
\usepackage{color}
\usepackage{ulem}

\usepackage{graphicx}
\usepackage{array}
\usepackage{arydshln}
\usepackage{tabularx}
\usepackage{multirow}
\usepackage{spverbatim}
\usepackage{todonotes}
\title{RecGPT: Generative Pre-training for Text-based Recommendation}

\author{Hoang Ngo \and Dat Quoc Nguyen  \\
         VinAI Research, Vietnam \\
        \{v.hoangnv49, v.datnq9\}@vinai.io}

\begin{document}
\maketitle
\begin{abstract}
We present the first domain-adapted and fully-trained large language model, RecGPT-7B, and its instruction-following variant, RecGPT-7B-Instruct, for text-based recommendation. Experimental results on rating prediction and sequential recommendation tasks show that our model, RecGPT-7B-Instruct, outperforms previous strong baselines. We are releasing our RecGPT models as well as their pre-training and fine-tuning datasets to facilitate future research and downstream applications in text-based recommendation. Public ``huggingface'' links to our RecGPT models and datasets are available at: \url{https://github.com/VinAIResearch/RecGPT}.
\end{abstract}

\section{Introduction}
Recommendation systems assist in comprehending user preferences and offering suitable content suggestions for users \cite{jmkr.37.3.363.18779,10.1145/352871.352887,Pazzani2007}.
Currently,  recommendation systems have found wide applications across various domains, such as e-commerce \cite{Schafer2001,sasrec}, news \cite{dkn}, and movies \cite{bert4rec}. 
The evolution of recommendation systems has witnessed a shift from fundamental methods to more sophisticated and modern approaches. Conventional methods mine interaction matrices to exploit user-item relationships \cite{mf, usenet, neumf}, and subsequently, they incorporate deep learning techniques such as CNN and RNN to extract item features and capture user preferences \cite{dkn, gru4rec}.
However, this task-specific setting suffers from data sparsity, a lack of flexibility to capture fluctuations in user preferences over time, and challenges in scaling to a large number of users and extensive datasets. Later works, inspired by attention mechanisms and the Transformers architecture \cite{transformers}, model user histories as sequences of items and then encode information in dense vectors \cite{sasrec, bert4rec, s3rec}.

With the advancement of large language models (LLMs), recent works leverage the capacity of LLMs in understanding user preferences \cite{geng-etal-2023-vip5,tiger}. The model P5 \cite{p5}, which represents users and items by IDs, endeavors to aggregate recommendation tasks under a unified conditional generation model based on T5 \cite{t5}. In addition, \citet{chatgptrec} evaluate the potential usage of ChatGPT in different recommendation tasks. More recently, \citet{genrec} fine-tune LLaMA \cite{touvron2023llama} with LoRA \cite{hu2022lora} for sequential recommendation. 
Recommendation tasks frequently exhibit shared characteristics such as user sets, item sets, and interactions, thus suggesting the possibility of training a unified model for multiple tasks, as opposed to employing distinct models for each task. Adopting a single model approach, as done in P5, not only encourages model generalization but also fosters collaborative learning across tasks. 
However, representing users and items by IDs, as in P5, may not fully align with the textual understanding capability of LLMs. It might be more effective to represent items by their textual descriptions and users by their text-based interaction history with items. 

In this paper, \textbf{(I)} we introduce the first domain-adapted and fully-trained LLM series named RecGPT for text-based recommendation, which comprises the base pre-trained model RecGPT-7B and its instruction-following variant, RecGPT-7B-Instruct. In this context, we pre-train RecGPT-7B using a relatively large recommendation-specific corpus of 20.5B tokens, while RecGPT-7B-Instruct is the model output by further fine-tuning RecGPT-7B on a dataset of 100K+ instructional prompts and their responses. \textbf{(II)} We conduct experiments for rating prediction and sequential recommendation tasks, demonstrating that our RecGPT-7B-Instruct outperforms strong baselines, including P5. \textbf{(III)} We publicly release our models along with the pre-training and fine-tuning datasets. We hope that this release can foster future research and applications in text-based recommendation.

\section{Our model RecGPT}

This section describes the data and outlines the architecture and optimization setup used for RecGPT.

\begin{table}[t!]
\scriptsize \centering
\setlength\tabcolsep{2pt}
\begin{tabularx}{\linewidth}{l|l}
\hline
\multicolumn{2}{l}{\textbf{Pre-training sample (showing the first 3 items for illustration)}} \\
\hline
\multicolumn{1}{l}{text} &
\begin{tabular}[c]{@{}X@{}}
Given the interaction history of a user with products as follows: \\
Title: Rock-a-Stack; Brand: Fisher-Price; Review: My son loves to empty this stacker and play with and teeth on the rings; Rating: 5.0/5.0 \\
Title: Jumbo Puzzle; Brand: Melissa \& Doug; Review: My niece love this puzzle at my parents house so I had to have it for my son. A classic!; Rating: 5.0/5.0 \\
Title: So Big Crayons; Brand: Crayola; Review: Good quality as expected from Crayola and easy enough for him to grasp.; Rating: 5.0/5.0 \\ ... \\
\end{tabular} \\
\hline
\hline
\multicolumn{2}{l}{\textbf{Fine-tuning samples}} \\
\hline
\multicolumn{1}{l}{prompt} &
\begin{tabular}[c]{@{}X@{}}
Predict the rating for the last item. Given the interaction history of a user with products as follows:\\
Title: Frankenweenie Figure; Brand: Disney; Review: My daughter loves Frankenweenie \& I was super excited to find Sparky on here; Rating: 5.0/5.0\\ Title: Rubber Ghost Face; Brand: Fun World; Review: The rubber is so flimsy it literally flaps in the wind when you move your hand while holding it. Rating: 2.0/5.0\\ Title: Makeup Signature Set; Brand: LCosmetics; Review: The rubber is so flimsy it literally flaps in the wind when you move your hand while holding it.; Rating: 4.0/5.0\\ Title: Hive Building Sets; Brand: HEXBUG; Review: It is fun \& my daughter loves it; Rating: 
\end{tabular} \\
\hdashline
\multicolumn{1}{l}{response} & 
\begin{tabular}[c]{@{}X@{}}
4.0/5.0
\end{tabular} \\
\hline
\multicolumn{1}{l}{prompt} &
\begin{tabular}[c]{@{}X@{}}
Predict the next item. Given the interaction history of a user with products as follows:\\
Title: Frankenweenie Figure; Brand: Disney\\ Title: Rubber Ghost Face; Brand: Fun World\\ Title: Makeup Signature Set; Brand: LCosmetics\\ Title: Hive Building Sets; Brand: HEXBUG
\end{tabular} \\
\hdashline
\multicolumn{1}{l}{response} & 
\begin{tabular}[c]{@{}X@{}}
Title: Animal Hats; Brand: ZoopurPets
\end{tabular} \\
\hline
\end{tabularx}
\caption{Pre-training and fine-tuning data examples.}
\label{tab:data-ex}
\end{table}

\subsection{Pre-training and Fine-tuning data}
\label{sec:method-data}

We collect a rich and comprehensive set of datasets from various domains, including: Amazon Product \cite{amazon}, Anime,\footnote{\tiny \url{https://www.kaggle.com/datasets/CooperUnion/anime-recommendations-database}} BookCrossing,\footnote{\tiny  \url{https://www.kaggle.com/datasets/ruchi798/bookcrossing-dataset}} Food \cite{food}, Goodreads \cite{goodreads}, HotelRec \cite{hotelrec}, MovieLens \cite{movielens}, Netflix \cite{netflix}, Steam,\footnote{\tiny  \url{https://www.kaggle.com/datasets/tamber/steam-video-games}} WikiRec  \cite{wikirec}, and Yelp.\footnote{\tiny  \url{https://www.yelp.com/dataset}} Specifically, we select datasets that contain item \textit{titles}, a key factor for item representation. 
Each item is associated with metadata comprising attributes such as \textit{title} and \textit{brand}, along with user interactions such as \textit{rating} and \textit{review}.   
We perform a cleaning pre-process on the collected datasets by discarding: (i) items without titles, (ii) users with fewer than 5 interactions, and (iii) all background and demographic user information. Ultimately, we have 10,156,309 users, 10,309,169 items, and 258,100,698 interactions in total. Detailed statistics of each cleaned dataset are shown in Table \ref{tab:data-train} in Appendix \ref{sec:apd-data}.

Then we randomly split each cleaned dataset into pre-training/fine-tuning subsets with a 99.5/0.5 ratio at the ``user'' level (i.e., users in the fine-tuning subset do not appear in the pre-training subset, and vice versa).\footnote{There are 4 datasets where we do not apply the 99.5/0.5 ratio. Refer to Section \ref{ssec:benchmarks} for more details.}  Regarding pre-training, users are represented solely through their interaction history with items. Each user's interaction history, referred to as a text document, is formatted as a chronologically-ordered list of text-based data points $i_1, i_{2}, ..., i_n$, where $i_k$ is represented by the corresponding $k$-th item's metadata and interactions. For example, in the pre-training sample in Table~\ref{tab:data-ex}, $i_1$ is ``\texttt{\small Title: Rock-a-Stack; Brand: Fisher-Price; Review: My son loves to empty this stacker and play with and teeth on the rings; Rating: 5.0/5.0}''. 
Totally, we create a pre-training corpus of 10M+ documents with 20.5B tokens. 

When it comes to fine-tuning for instruction following, given the nature of our datasets, we create prompt-response pairs for two popular tasks in the recommendation system domain: \textit{rating prediction} and \textit{sequential recommendation}. For each user with the history $i_1, i_{2}, ..., i_n$, the last item $i_n$ is considered as the next item to be predicted in sequential recommendation, given the history context $i_1, i_{2}, ..., i_{n-1}$. Meanwhile, the rating of the $(n-1)$-th item $i_{n-1}$ is used as the label for rating prediction, given the remaining history context $i_1, i_{2}, ..., i_{n-1}$ without the rating of the $(n-1)$-th item. Depending on task requirements, unused features within each data point $i_k$ of the user history are discarded, streamlining the prompts and their responses for enhanced task relevance and efficiency. Altogether, we create a fine-tuning dataset of 100K+ instructional prompt and response pairs.

Examples of a pre-training document and prompt-response pairs are shown in Table~\ref{tab:data-ex}. Details on the data formats used in pre-training and fine-tuning are presented in Appendix \ref{sec:apd-prompt}.

\subsection{RecGPT-7B}


RecGPT-7B is a Transformer decoder-based model \citep{NEURIPS2020_1457c0d6,NIPS2017_7181} that incorporates (Triton) flash attention \citep{dao2022flashattention} and ALiBi \citep{press2022train} for context length extrapolation. Additionally, we use a ``max\_seq\_len'' of 2048, ``d\_model'' of 4096, ``n\_heads'' of 32, ``n\_layers'' of 32, and GPT-NeoX's tokenizer with a vocabulary of 50K tokens, resulting in a model size of about 7B parameters. Utilizing the Mosaicml ``llm-foundry'' library,\footnote{\url{https://github.com/mosaicml/llm-foundry}: A robust library that supports both pre-training and fine-tuning.} 
we initialize the parameter weights of RecGPT-7B with those from the pre-trained MPT-7B \citep{MosaicML2023Introducing} and continually pre-train on our pre-training corpus of 20.5B tokens. For optimization, we employ the LION optimizer \citep{chen2023symbolic}  and sharded data parallelism with FSDP, set a global batch size of 128 (i.e., 128 * 2048 = 260K tokens per batch) across 8 A100 GPUs (40GB each), and use a peak learning rate of 2.5e-5. The training runs for 2 epochs, using mixed precision training with bfloat16, and takes about 18 days. This is equivalent to 20.5B * 2 / 260K = 157K training steps (here, the learning rate is warmed up for the first 2K training steps).

The total number of GPU hours used for pre-training is 18 * 8 * 24 = 3456. With the GPU power consumption at 400W, the pre-training process uses 3456 * 400 = 1,382,400 Wh, equivalent to the carbon emission of about 0.585 tCO2eq.

\subsection{RecGPT-7B-Instruct}\label{ssec:instructionfollowing}

We then fine-tune the base pre-trained RecGPT-7B for instruction following regarding rating prediction and sequential recommendation, using the dataset consisting of 100K+ instructional prompts and their responses from Section \ref{sec:method-data}. We employ LION, set a global batch size of 128 across 8 A100 GPUs (40GB each), use a peak learning rate of 1.0e-5, and run for 2 epochs. The resulting fine-tuned model is named RecGPT-7B-Instruct.

Fine-tuning RecGPT-7B-Instruct takes 4 hours using a node of 8 A100 GPUs (40GB each), totaling 32 GPU hours. This is equivalent to the carbon emission of about 0.0054 tCO2eq.

\begin{table*}[t!]
\centering
\small
\begin{tabular}{l|rr|rr|rr|rr}
\hline
& \multicolumn{2}{c|}{\textbf{Beauty}} & \multicolumn{2}{c|}{\textbf{Sport}} & \multicolumn{2}{c|}{\textbf{Toys}} & \multicolumn{2}{c}{\textbf{Yelp}} \\
\textbf{Model} & \textbf{RMSE} & \textbf{MAE} & \textbf{RMSE} & \textbf{MAE} & \textbf{RMSE} & \textbf{MAE} & \textbf{RMSE} & \textbf{MAE} \\ \hline
MF \cite{mf} [*] & 1.1973 & 0.9461 & 1.0234 & 0.7935 & 1.0123 & 0.7984 & 1.2645 & 1.0426 \\
MLP \cite{mlp} [*] & 1.3078 & 0.9597 & 1.1277 & 0.7626 & 1.1215 & 0.8097 & 1.2951 & 1.0340 \\
P5 \cite{p5} [*] & 1.2843 & 0.8534 & 1.0357 & 0.6813 & 1.0544 & 0.7177 & 1.4685 & 1.0054\\ 
ChatGPT (few-shot) [$\dagger$] & 1.0751 & 0.6977 & - & - & - & - & - & - \\ \hdashline
MPT-7B with SFT & 0.5637 & 0.2616 & 0.5446 & 0.2488 & 0.5565 & 0.2668 & 0.5620 & 0.2804 \\
RecGPT-7B-Instruct & \textbf{0.5316} & \textbf{0.2436} & \textbf{0.5208} & \textbf{0.2340} & \textbf{0.5361} & \textbf{0.2535} & \textbf{0.5203} & \textbf{0.2489} \\ \hline
\end{tabular}
\caption{Results obtained for rating prediction: ``Sport'' and ``Toys'' abbreviate ``Sports and Outdoors'' and ``Toys and Games'', respectively. [*] denotes results reported by \citet{p5}. [$\dagger$] denotes the results of the best model ChatGPT (GPT-3.5-turbo) among different models experimented with by \citet{chatgptrec}.}
\label{tab:res-rating}
\end{table*}

\section{Experiments}

We conduct experiments to compare our RecGPT-7B-Instruct with strong baselines for rating prediction and sequential recommendation tasks.

\subsection{Experimental setup}\label{ssec:benchmarks}

\paragraph{Evaluation datasets:} We carry out experiments on 4 benchmark datasets across different domains, including  ``Amazon Beauty'', ``Amazon Sports and Outdoors'' and ``Amazon Toys and Games'' \cite{amazon}, as well as Yelp. Following previous works \citep{p5, genrec}, for those three Amazon datasets, we employ the 5-core version 2014,\footnote{\url{https://cseweb.ucsd.edu/~jmcauley/datasets/amazon/links.html}} while for Yelp, we consider transactions from Jan 1, 2019, to Dec 31, 2019.

\underline{Data leakage issue:} We further discover a data leakage issue that has not been pointed out before. As the four experimental benchmark datasets used in the evaluation are not pre-defined with a training-validation-test split, previous works apply different splitting strategies for each evaluation task \cite{p5}. 
Let's consider the Amazon Beauty dataset, which is utilized in training P5 \cite{p5}, as an example (similar findings apply to other datasets). The dataset comprises users, items, and interactions between them. An interaction example may be: user X purchasing item Y and providing a review and rating of 4.0/5.0. The original dataset is presented as interaction records without a predefined training-validation-test split.
P5 employs different data splitting strategies for different tasks. For the rating prediction task, P5 randomly divides the data into training, validation, and test sets with an 80-10-10 ratio, respectively. For the sequential recommendation task, P5 aggregates data by user to construct users' histories, comprising their interactions. Then, P5 utilizes a leave-one-out manner, where the last item in the history is reserved for testing, the second-last item for validation, and the remaining items for training. Consequently, there are interactions in the training set for the rating prediction task, which also belong to the test set for the sequential recommendation task, and vice versa (i.e., there are interactions in the training set in the sequential recommendation task, which also belong to the test set in the rating prediction task). Merging the training sets from both tasks for multitask training, as performed in P5, without filtering out duplicate data results in data leakage. 

For a consistent test set, we still reuse their splits but remove interactions from the training set if they appear in the test set. This ensures that the test data is not leaked into the training data. Note that for these 4 experimental benchmarks, we report our final scores on the test split, while the training split is only used for pre-training RecGPT-7B to mimic real-world scenarios (i.e., we do not use the training/validation split for supervised fine-tuning of instruction following).

\paragraph{Evaluation metrics:} For rating prediction, we employ Root Mean Square Error (RMSE) and Mean Absolute Error (MAE), while for sequential recommendation, we use top-k Hit Ratio (HR@k) and top-k Normalized Discounted Cumulative Gain (NDCG@k). Smaller values of RMSE and MAE, and higher values of HR and NDCG, indicate better performance.

\paragraph{Inference:} 
We utilize vLLM \cite{vllm} as an inference engine. For rating prediction, for a given input prompt,  we apply the sampling decoding strategy with ``temperature'' of 1.0, ``top\_p'' of 0.9 and ``top\_k'' set at 50, and then extract the predicted value from the generated response output. For sequential recommendation, following previous works \citep{p5,genrec}, for a given input prompt,  we use the beam search decoding strategy with a beam size of 10 to generate 10 response outputs and use their beam search scores for ranking. In addition, due to the hallucinatory nature of LLMs, the generated outputs might differ slightly from the ground truth labels. Therefore, we implement a semantic similarity matching approach with a text embedding model and a matching module, built on top of Sentence Transformers \cite{sentence-transformers} and FAISS \cite{faiss} respectively. This approach utilizes dot product-based similarity over dense vector representations to associate each generated output with the most similar item in the item set.

\subsection{Main results}

\paragraph{Rating prediction:} Table~\ref{tab:res-rating} lists rating prediction results for our RecGPT-7B-Instruct and the previous strong baselines on the four experimental datasets. We find that, in general, pre-trained LLM-based approaches, specifically P5 \cite{p5}, ChatGPT (GPT-3.5-turbo), and  RecGPT-7B-Instruct, outperform conventional rating prediction methods MF \cite{mf} and MLP \cite{mlp}. Although ChatGPT is not specifically designed for this task, it demonstrates promising performance scores that surpass those of P5 on the ``Beauty'' dataset. We find that RecGPT-7B-Instruct achieves the best results across all datasets in terms of both evaluation metrics  RMSE and MAE, yielding new state-of-the-art performance scores.

\begin{table}[t!]
\centering
\small
\setlength\tabcolsep{2pt} 
\begin{tabular}{ll|r|r|r|r}
\hline
& \textbf{Model} &
  \textbf{\begin{tabular}[c]{@{}r@{}}HR\\ @5\end{tabular}} &
  \textbf{\begin{tabular}[c]{@{}r@{}}NDCG\\ @5\end{tabular}} &
  \textbf{\begin{tabular}[c]{@{}r@{}}HR\\ @10\end{tabular}} &
  \textbf{\begin{tabular}[c]{@{}r@{}}NDCG\\ @10\end{tabular}} \\ \hline
\multirow{5}{*}{\rotatebox[origin=c]{90}{\textbf{Beauty}}} & P5 [$\star$]  & 0.0350 &  \textbf{0.0250} &  0.0480 &  \textbf{0.0298} \\
& ChatGPT (few-shot) ($\dagger$)& 0.0135 & 0.0135 & 0.0135 & 0.0135 \\
& OpenP5 (\citeauthor{openp5}) & 0.0317 & {0.0239} & 0.0437 & {0.0277}  \\
\cdashline{2-6}
& MPT-7B with SFT  & 0.0063 & 0.0041 & 0.0088 & 0.0050 \\
& RecGPT-7B-Instruct & \textbf{0.0364} & 0.0236 & \textbf{0.0527} & 0.0288 \\ \hline
\hline
\multirow{4}{*}{\rotatebox[origin=c]{90}{\textbf{Toys}}} 
& P5 [$\star$] &  0.0180 & 0.0130 & 0.0235 & 0.0150        \\
& GenRec (\citeauthor{genrec}) & 0.0190 & 0.0136 & 0.0251 & 0.0157 \\ 
\cdashline{2-6}
& MPT-7B with SFT  & 0.0088 & 0.0061 & 0.0133 & 0.0075 \\
& RecGPT-7B-Instruct & \textbf{0.0430} & \textbf{0.0288} & \textbf{0.0606} & \textbf{0.0343} \\ 
\hline
\hline
\multirow{3}{*}{\rotatebox[origin=c]{90}{\textbf{Sport}}} & P5 [$\star$] & 0.0107 & 0.0076 & 0.0146 & 0.0088 \\
\cdashline{2-6}
& MPT-7B with SFT    & 0.0021 & 0.0015 & 0.0033 & 0.0018  \\
& RecGPT-7B-Instruct & \textbf{0.0173} & \textbf{0.0110} & \textbf{0.0255} & \textbf{0.0136} \\
\hline
\hline
\multirow{2}{*}{\rotatebox[origin=c]{90}{\textbf{Yelp}}} 
& MPT-7B with SFT     & 0.0390 & 0.0280 & 0.0453 & 0.0298 \\
& RecGPT-7B-Instruct & \textbf{0.0479} & \textbf{0.0339} & \textbf{0.0603} & \textbf{0.0377} \\
\hline
\end{tabular}
\caption{Results obtained for sequential recommendation. [$\star$] denotes P5's results with standard pre-processing, as reported by \citet{tiger}, where they do not conduct experiments on the Yelp dataset.}
\label{tab:res-seq}
\end{table}

\paragraph{Sequential recommendation:} Table~\ref{tab:res-seq} presents the obtained results with cutoff thresholds of 5 and 10 for HR and NDCG for different models on the sequential recommendation task. 
Not surprisingly, ChatGPT, which faces a limitation in terms of in-domain data, attains lower scores than other baselines on the ``Beauty'' dataset. This highlights the crucial role of in-domain training data in sequential recommendation for models to comprehend the item set. 
GenRec \citep{genrec}, fine-tuned with LoRa \citep{hu2022lora} on the entire training split, does not perform competitively on the ``Toys and Games'' dataset, compared to the fully fine-tuned model RecGPT-7B-Instruct. Additionally, our RecGPT-7B-Instruct achieves competitive results with P5 and OpenP5 \citep{openp5} on the ``Beauty'' dataset. Moreover, RecGPT-7B-Instruct notably outperforms P5 on both the ``Sports and Outdoors'' and ``Toys and Games'' datasets.

\paragraph{Ablation analysis:} To examine how pre-training contributes to the improvement in the performance scores of RecGPT-7B-Instruct, we also conduct supervised fine-tuning (SFT) for instruction following on the base pre-trained MPT-7B. The fine-tuning process for MPT-7B is carried out in the same manner as for our RecGPT-7B-Instruct, as detailed in Section \ref{ssec:instructionfollowing}. Tables \ref{tab:res-rating} and \ref{tab:res-seq} also present the results of MPT-7B with SFT. We find that RecGPT-7B-Instruct performs substantially better than MPT-7B with SFT, highlighting the significant contribution of continual pre-training RecGPT-7B for domain adaptation in the context of recommendation.

In Table~\ref{tab:res-rating}, rating prediction most likely relies on the review text to predict the score, which might be viewed as a sentiment classification task with more fine-grained labels. This task is thus not as difficult (compared to the sequential recommendation task), given tens of thousands of examples for rating prediction fine-tuning. Also, the base LLM model MPT-7B is pre-trained on a 1T-token corpus that likely contains many reviews from the web. So the substantial improvement of RecGPT-7B-Instruct over the baseline ``MPT-7B with SFT'' for the rating prediction task is not as large as for the sequential recommendation task.

\section{Conclusion}

We have introduced the first domain-adapted and fully-trained LLMs for text-based recommendation, which include the base pre-trained RecGPT-7B and its instruction-following variant, RecGPT-7B-Instruct. We demonstrate the usefulness of RecGPT by showing that RecGPT-7B-Instruct outperforms strong baselines in both rating prediction and sequential recommendation tasks. Through the public release of RecGPT models and the pre-training and supervised fine-tuning datasets, we hope that they can foster future research and applications in text-based recommendation.

\section*{Limitations}
The knowledge of the LLM about the tasks and the item set is solely based on training data and the intrinsic memory of the base model. Models might not be aware of items that are not covered in the training data. If this incident occurs, models could generate irrelevant information and suffer from hallucinations. This limitation also applies to all LLM-based methods. Furthermore, in this work, we only evaluate two popular tasks; we will conduct experiments for other recommendation tasks in future work.

\section*{Acknowledgement}

We extend our thanks to Khoa D. Doan (khoa.dd@vinuni.edu.vn) for initial discussions.

\bibliography{custom}

\begin{thebibliography}{39}
\expandafter\ifx\csname natexlab\endcsname\relax\def\natexlab#1{#1}\fi

\bibitem[{AlGhamdi et~al.(2021)AlGhamdi, Shi, and Simperl}]{wikirec}
Kholoud AlGhamdi, Miaojing Shi, and Elena Simperl. 2021.
\newblock {Learning to Recommend Items to Wikidata Editors}.
\newblock In \emph{The Semantic Web – ISWC 2021: 20th International Semantic Web Conference, ISWC 2021, Virtual Event, October 24–28, 2021, Proceedings}, page 163–181.

\bibitem[{Ansari et~al.(2000)Ansari, Essegaier, and Kohli}]{jmkr.37.3.363.18779}
Asim Ansari, Skander Essegaier, and Rajeev Kohli. 2000.
\newblock {Internet Recommendation Systems}.
\newblock \emph{Journal of Marketing Research}, 37(3):363--375.

\bibitem[{Antognini and Faltings(2020)}]{hotelrec}
Diego Antognini and Boi Faltings. 2020.
\newblock {HotelRec: a Novel Very Large-Scale Hotel Recommendation Dataset}.
\newblock In \emph{Proceedings of the Twelfth Language Resources and Evaluation Conference}, pages 4917--4923.

\bibitem[{Bennett and Lanning(2007)}]{netflix}
James Bennett and Stan Lanning. 2007.
\newblock {The Netflix Prize}.
\newblock In \emph{Proceedings of KDD Cup and Workshop 2007}, page~35.

\bibitem[{Brown et~al.(2020)Brown, Mann, Ryder, Subbiah, Kaplan, Dhariwal, Neelakantan, Shyam, Sastry, Askell, Agarwal, Herbert-Voss, Krueger, Henighan, Child, Ramesh, Ziegler, Wu, Winter, Hesse, Chen, Sigler, Litwin, Gray, Chess, Clark, Berner, McCandlish, Radford, Sutskever, and Amodei}]{NEURIPS2020_1457c0d6}
Tom Brown, Benjamin Mann, Nick Ryder, Melanie Subbiah, Jared~D Kaplan, Prafulla Dhariwal, Arvind Neelakantan, Pranav Shyam, Girish Sastry, Amanda Askell, Sandhini Agarwal, Ariel Herbert-Voss, Gretchen Krueger, Tom Henighan, Rewon Child, Aditya Ramesh, Daniel Ziegler, Jeffrey Wu, Clemens Winter, Chris Hesse, Mark Chen, Eric Sigler, Mateusz Litwin, Scott Gray, Benjamin Chess, Jack Clark, Christopher Berner, Sam McCandlish, Alec Radford, Ilya Sutskever, and Dario Amodei. 2020.
\newblock {Language Models are Few-Shot Learners}.
\newblock In \emph{Proceedings of NeurIPS}.

\bibitem[{Chen et~al.(2023)Chen, Liang, Huang, Real, Wang, Pham, Dong, Luong, Hsieh, Lu, and Le}]{chen2023symbolic}
Xiangning Chen, Chen Liang, Da~Huang, Esteban Real, Kaiyuan Wang, Hieu Pham, Xuanyi Dong, Thang Luong, Cho-Jui Hsieh, Yifeng Lu, and Quoc~V Le. 2023.
\newblock Symbolic discovery of optimization algorithms.
\newblock In \emph{Thirty-seventh Conference on Neural Information Processing Systems}.

\bibitem[{Cheng et~al.(2016)Cheng, Koc, Harmsen, Shaked, Chandra, Aradhye, Anderson, Corrado, Chai, Ispir, Anil, Haque, Hong, Jain, Liu, and Shah}]{mlp}
Heng-Tze Cheng, Levent Koc, Jeremiah Harmsen, Tal Shaked, Tushar Chandra, Hrishi Aradhye, Glen Anderson, Greg Corrado, Wei Chai, Mustafa Ispir, Rohan Anil, Zakaria Haque, Lichan Hong, Vihan Jain, Xiaobing Liu, and Hemal Shah. 2016.
\newblock {Wide \& Deep Learning for Recommender Systems}.
\newblock In \emph{Proceedings of the 1st Workshop on Deep Learning for Recommender Systems}, page 7–10.

\bibitem[{Dao et~al.(2022)Dao, Fu, Ermon, Rudra, and R{\'e}}]{dao2022flashattention}
Tri Dao, Daniel~Y. Fu, Stefano Ermon, Atri Rudra, and Christopher R{\'e}. 2022.
\newblock {Flash{A}ttention: Fast and Memory-Efficient Exact Attention with {IO}-Awareness}.
\newblock In \emph{Proceedings of NeurIPS}.

\bibitem[{Geng et~al.(2022)Geng, Liu, Fu, Ge, and Zhang}]{p5}
Shijie Geng, Shuchang Liu, Zuohui Fu, Yingqiang Ge, and Yongfeng Zhang. 2022.
\newblock {Recommendation as Language Processing (RLP): A Unified Pretrain, Personalized Prompt \& Predict Paradigm (P5)}.
\newblock In \emph{Proceedings of the 16th ACM Conference on Recommender Systems}, page 299–315.

\bibitem[{Geng et~al.(2023)Geng, Tan, Liu, Fu, and Zhang}]{geng-etal-2023-vip5}
Shijie Geng, Juntao Tan, Shuchang Liu, Zuohui Fu, and Yongfeng Zhang. 2023.
\newblock {VIP}5: Towards multimodal foundation models for recommendation.
\newblock In \emph{Findings of the Association for Computational Linguistics: EMNLP 2023}, pages 9606--9620.

\bibitem[{Harper and Konstan(2015)}]{movielens}
F.~Maxwell Harper and Joseph~A. Konstan. 2015.
\newblock {The MovieLens Datasets: History and Context}.
\newblock \emph{ACM Trans. Interact. Intell. Syst.}, 5(4):1--19.

\bibitem[{He et~al.(2017)He, Liao, Zhang, Nie, Hu, and Chua}]{neumf}
Xiangnan He, Lizi Liao, Hanwang Zhang, Liqiang Nie, Xia Hu, and Tat-Seng Chua. 2017.
\newblock {Neural Collaborative Filtering}.
\newblock In \emph{Proceedings of the 26th International Conference on World Wide Web}, page 173–182.

\bibitem[{Hidasi et~al.(2016)Hidasi, Karatzoglou, Baltrunas, and Tikk}]{gru4rec}
Bal{\'{a}}zs Hidasi, Alexandros Karatzoglou, Linas Baltrunas, and Domonkos Tikk. 2016.
\newblock {Session-based Recommendations with Recurrent Neural Networks}.
\newblock In \emph{4th International Conference on Learning Representations, {ICLR} 2016, San Juan, Puerto Rico, May 2-4, 2016, Conference Track Proceedings}.

\bibitem[{Hu et~al.(2022)Hu, yelong shen, Wallis, Allen-Zhu, Li, Wang, Wang, and Chen}]{hu2022lora}
Edward~J Hu, yelong shen, Phillip Wallis, Zeyuan Allen-Zhu, Yuanzhi Li, Shean Wang, Lu~Wang, and Weizhu Chen. 2022.
\newblock \href {https://openreview.net/forum?id=nZeVKeeFYf9} {Lo{RA}: Low-rank adaptation of large language models}.
\newblock In \emph{International Conference on Learning Representations}.

\bibitem[{Ji et~al.(2024)Ji, Li, Xu, Hua, Ge, Tan, and Zhang}]{genrec}
Jianchao Ji, Zelong Li, Shuyuan Xu, Wenyue Hua, Yingqiang Ge, Juntao Tan, and Yongfeng Zhang. 2024.
\newblock {GenRec: Large Language Model for Generative Recommendation}.
\newblock In \emph{Proceedings of the 46th European Conference on Information Retrieval}, page to appear.

\bibitem[{Johnson et~al.(2021)Johnson, Douze, and Jégou}]{faiss}
Jeff Johnson, Matthijs Douze, and Hervé Jégou. 2021.
\newblock {Billion-Scale Similarity Search with GPUs}.
\newblock \emph{IEEE Transactions on Big Data}, pages 535--547.

\bibitem[{Kang and McAuley(2018)}]{sasrec}
Wang-Cheng Kang and Julian McAuley. 2018.
\newblock {Self-Attentive Sequential Recommendation}.
\newblock In \emph{2018 IEEE International Conference on Data Mining (ICDM)}, pages 197--206.

\bibitem[{Konstan et~al.(1997)Konstan, Miller, Maltz, Herlocker, Gordon, and Riedl}]{usenet}
Joseph~A. Konstan, Bradley~N. Miller, David Maltz, Jonathan~L. Herlocker, Lee~R. Gordon, and John Riedl. 1997.
\newblock {GroupLens: applying collaborative filtering to Usenet news}.
\newblock \emph{Commun. ACM}, page 77–87.

\bibitem[{Koren et~al.(2009)Koren, Bell, and Volinsky}]{mf}
Yehuda Koren, Robert Bell, and Chris Volinsky. 2009.
\newblock {Matrix Factorization Techniques for Recommender Systems}.
\newblock \emph{Computer}, 42:30--37.

\bibitem[{Kwon et~al.(2023)Kwon, Li, Zhuang, Sheng, Zheng, Yu, Gonzalez, Zhang, and Stoica}]{vllm}
Woosuk Kwon, Zhuohan Li, Siyuan Zhuang, Ying Sheng, Lianmin Zheng, Cody~Hao Yu, Joseph~E. Gonzalez, Hao Zhang, and Ion Stoica. 2023.
\newblock {Efficient Memory Management for Large Language Model Serving with PagedAttention}.
\newblock In \emph{Proceedings of the ACM SIGOPS 29th Symposium on Operating Systems Principles}.

\bibitem[{Liu et~al.(2023)Liu, Liu, Zhou, Lv, Zhou, and Zhang}]{chatgptrec}
Junling Liu, Chao Liu, Peilin Zhou, Renjie Lv, Kang Zhou, and Yan Zhang. 2023.
\newblock {Is ChatGPT a Good Recommender? A Preliminary Study}.
\newblock In \emph{Proceedings of the the 1st CIKM Workshop on Recommendation with Generative Models}.

\bibitem[{Majumder et~al.(2019)Majumder, Li, Ni, and McAuley}]{food}
Bodhisattwa~Prasad Majumder, Shuyang Li, Jianmo Ni, and Julian McAuley. 2019.
\newblock {Generating Personalized Recipes from Historical User Preferences}.
\newblock In \emph{Proceedings of the 2019 Conference on Empirical Methods in Natural Language Processing and the 9th International Joint Conference on Natural Language Processing}, pages 5976--5982.

\bibitem[{McAuley et~al.(2015)McAuley, Targett, Shi, and van~den Hengel}]{amazon}
Julian McAuley, Christopher Targett, Qinfeng Shi, and Anton van~den Hengel. 2015.
\newblock {Image-Based Recommendations on Styles and Substitutes}.
\newblock In \emph{Proceedings of the 38th International ACM SIGIR Conference on Research and Development in Information Retrieval}, page 43–52.

\bibitem[{Pazzani and Billsus(2007)}]{Pazzani2007}
Michael~J. Pazzani and Daniel Billsus. 2007.
\newblock \emph{{Content-Based Recommendation Systems}}, pages 325--341.

\bibitem[{Press et~al.(2022)Press, Smith, and Lewis}]{press2022train}
Ofir Press, Noah Smith, and Mike Lewis. 2022.
\newblock {Train Short, Test Long: Attention with Linear Biases Enables Input Length Extrapolation}.
\newblock In \emph{Proceedings of ICLR}.

\bibitem[{Raffel et~al.(2020)Raffel, Shazeer, Roberts, Lee, Narang, Matena, Zhou, Li, and Liu}]{t5}
Colin Raffel, Noam Shazeer, Adam Roberts, Katherine Lee, Sharan Narang, Michael Matena, Yanqi Zhou, Wei Li, and Peter~J. Liu. 2020.
\newblock {Exploring the limits of transfer learning with a unified text-to-text transformer}.
\newblock \emph{J. Mach. Learn. Res.}

\bibitem[{Rajput et~al.(2023)Rajput, Mehta, Singh, Keshavan, Vu, Heldt, Hong, Tay, Tran, Samost, Kula, Chi, and Sathiamoorthy}]{tiger}
Shashank Rajput, Nikhil Mehta, Anima Singh, Raghunandan~Hulikal Keshavan, Trung Vu, Lukasz Heldt, Lichan Hong, Yi~Tay, Vinh~Q. Tran, Jonah Samost, Maciej Kula, Ed~H. Chi, and Maheswaran Sathiamoorthy. 2023.
\newblock {Recommender Systems with Generative Retrieval}.
\newblock In \emph{Proceedings of the Thirty-seventh Conference on Neural Information Processing Systems}.

\bibitem[{Reimers and Gurevych(2019)}]{sentence-transformers}
Nils Reimers and Iryna Gurevych. 2019.
\newblock {Sentence-BERT: Sentence Embeddings using {S}iamese {BERT}-Networks}.
\newblock In \emph{Proceedings of the 2019 Conference on Empirical Methods in Natural Language Processing and the 9th International Joint Conference on Natural Language Processing}, pages 3982--3992.

\bibitem[{Sarwar et~al.(2000)Sarwar, Karypis, Konstan, and Riedl}]{10.1145/352871.352887}
Badrul Sarwar, George Karypis, Joseph Konstan, and John Riedl. 2000.
\newblock Analysis of recommendation algorithms for e-commerce.
\newblock In \emph{Proceedings of the 2nd ACM Conference on Electronic Commerce}, page 158–167.

\bibitem[{Schafer et~al.(2001)Schafer, Konstan, and Riedl}]{Schafer2001}
J~Ben Schafer, Joseph~A Konstan, and John Riedl. 2001.
\newblock {E-Commerce Recommendation Applications}.
\newblock \emph{Data Mining and Knowledge Discovery}, 5(1):115--153.

\bibitem[{Sun et~al.(2019)Sun, Liu, Wu, Pei, Lin, Ou, and Jiang}]{bert4rec}
Fei Sun, Jun Liu, Jian Wu, Changhua Pei, Xiao Lin, Wenwu Ou, and Peng Jiang. 2019.
\newblock {BERT4Rec: Sequential Recommendation with Bidirectional Encoder Representations from Transformer}.
\newblock In \emph{Proceedings of the 28th ACM International Conference on Information and Knowledge Management}, page 1441–1450.

\bibitem[{Team(2023)}]{MosaicML2023Introducing}
MosaicML~NLP Team. 2023.
\newblock \href {www.mosaicml.com/blog/mpt-7b} {{Introducing MPT-7B: A New Standard for Open-Source, Commercially Usable LLMs}}.

\bibitem[{Touvron et~al.(2023)Touvron, Lavril, Izacard, Martinet, Lachaux, Lacroix, Rozière, Goyal, Hambro, Azhar, Rodriguez, Joulin, Grave, and Lample}]{touvron2023llama}
Hugo Touvron, Thibaut Lavril, Gautier Izacard, Xavier Martinet, Marie-Anne Lachaux, Timothée Lacroix, Baptiste Rozière, Naman Goyal, Eric Hambro, Faisal Azhar, Aurelien Rodriguez, Armand Joulin, Edouard Grave, and Guillaume Lample. 2023.
\newblock {LLaMA: Open and Efficient Foundation Language Models}.
\newblock \emph{arXiv preprint}, arXiv:2302.13971.

\bibitem[{Vaswani et~al.(2017{\natexlab{a}})Vaswani, Shazeer, Parmar, Uszkoreit, Jones, Gomez, Kaiser, and Polosukhin}]{transformers}
Ashish Vaswani, Noam Shazeer, Niki Parmar, Jakob Uszkoreit, Llion Jones, Aidan~N Gomez, \L~ukasz Kaiser, and Illia Polosukhin. 2017{\natexlab{a}}.
\newblock {Attention is All you Need}.
\newblock In \emph{Advances in Neural Information Processing Systems}.

\bibitem[{Vaswani et~al.(2017{\natexlab{b}})Vaswani, Shazeer, Parmar, Uszkoreit, Jones, Gomez, Kaiser, and Polosukhin}]{NIPS2017_7181}
Ashish Vaswani, Noam Shazeer, Niki Parmar, Jakob Uszkoreit, Llion Jones, Aidan~N Gomez, {\L}ukasz Kaiser, and Illia Polosukhin. 2017{\natexlab{b}}.
\newblock {Attention is All you Need}.
\newblock In \emph{Proceedings of NIPS}, pages 5998--6008.

\bibitem[{Wan and McAuley(2018)}]{goodreads}
Mengting Wan and Julian~J. McAuley. 2018.
\newblock {Item recommendation on monotonic behavior chains}.
\newblock In \emph{Proceedings of the 12th {ACM} Conference on Recommender Systems}, pages 86--94.

\bibitem[{Wang et~al.(2018)Wang, Zhang, Xie, and Guo}]{dkn}
Hongwei Wang, Fuzheng Zhang, Xing Xie, and Minyi Guo. 2018.
\newblock {DKN: Deep Knowledge-Aware Network for News Recommendation}.
\newblock In \emph{Proceedings of the 2018 World Wide Web Conference}, page 1835–1844.

\bibitem[{Xu et~al.(2023)Xu, Hua, and Zhang}]{openp5}
Shuyuan Xu, Wenyue Hua, and Yongfeng Zhang. 2023.
\newblock {OpenP5: Benchmarking Foundation Models for Recommendation}.
\newblock \emph{arXiv preprint}, arXiv:2306.11134.

\bibitem[{Zhou et~al.(2020)Zhou, Wang, Zhao, Zhu, Wang, Zhang, Wang, and Wen}]{s3rec}
Kun Zhou, Hui Wang, Wayne~Xin Zhao, Yutao Zhu, Sirui Wang, Fuzheng Zhang, Zhongyuan Wang, and Ji-Rong Wen. 2020.
\newblock {S3-Rec: Self-Supervised Learning for Sequential Recommendation with Mutual Information Maximization}.
\newblock In \emph{Proceedings of the 29th ACM International Conference on Information \& Knowledge Management}, page 1893–1902.

\end{thebibliography}

\appendix
\section{Datasets}
\label{sec:apd-data}
The statistics of our cleaned datasets are presented in Table \ref{tab:data-train}. Note that some datasets have two versions associated with different publication times (e.g., Amazon and Yelp). To maintain consistent test data with previous works \citep{p5, openp5, chatgptrec}, we retain the older versions (2014 for Amazon and 2020 for Yelp) for testing purposes and use the newer versions (2018 for Amazon and 2021 for Yelp) to enrich our pre-training data. We filter out overlapped users along with their interactions in the newer dataset to prevent duplication and data leakage.

Note that if a user has a long interaction history with many items (i.e., the number of tokens exceeds the max\_seq\_length of 2048), we pre-split the history into smaller chunks with a similar number of items, ensuring that the number of tokens in each chunk is smaller than 2048. Each chunk is then considered a separate user's interaction history.

\begin{table*}[htp]
\small
\centering
\begin{tabular}{lrrr}
\hline
\textbf{Dataset} & \textbf{\# Users} & \textbf{\# Items} & \textbf{\# Interactions} \\ \hline
Amazon All Beauty (2018)           & 195        & 85         & 1,026       \\
Amazon AMAZON FASHION              & 377        & 31         & 2,985       \\
Amazon Appliances                  & 20         & 47         & 119         \\
Amazon Arts Crafts and Sewing      & 46,651     & 22,855     & 401,244     \\
Amazon Automotive                  & 181,146    & 79,315     & 1,576,030   \\
Amazon Books                       & 1,847,930  & 703,927    & 26,751,568  \\
Amazon CDs and Vinyl               & 95,287     & 67,599     & 1,193,065   \\
Amazon Cell Phones and Accessories & 155,665    & 48,172     & 1,105,606   \\
Amazon Clothing Shoes and Jewelry  & 1,167,022  & 376,853    & 10,628,886  \\
Amazon Digital Music               & 34         & 183        & 248         \\
Amazon Electronics                 & 696,614    & 159,934    & 6,346,560   \\
Amazon Gift Cards                  & 456        & 148        & 2,961       \\
Amazon Grocery and Gourmet Food    & 116,141    & 41,280     & 1,024,096   \\
Amazon Home and Kitchen            & 733,886    & 189,038    & 6,406,439   \\
Amazon Industrial and Scientific   & 9,391      & 5,327      & 66,091      \\
Amazon Kindle Store                & 138,030    & 98,118     & 2,178,518   \\
Amazon Luxury Beauty               & 2,779      & 1,577      & 25,386      \\
Amazon Magazine Subscriptions      & 309        & 151        & 2,120       \\
Amazon Movies and TV               & 282,072    & 60,109     & 3,199,604   \\
Amazon Musical Instruments         & 25,402     & 10,611     & 210,646     \\
Amazon Office Products             & 88,788     & 27,931     & 689,303     \\
Amazon Patio Lawn and Garden       & 91,297     & 32,869     & 694,084     \\
Amazon Pet Supplies                & 213,455    & 42,498     & 1,854,600   \\
Amazon Prime Pantry                & 13,139     & 4,968      & 127,351     \\
Amazon Software                    & 1,470      & 802        & 10,571      \\
Amazon Sports and Outdoors (2018)  & 302,870    & 104,559    & 2,541,948   \\
Amazon Tools and Home Improvement  & 220,804    & 73,548     & 1,865,844   \\
Amazon Toys and Games (2018)       & 194,141    & 78,695     & 1,687,243   \\
Amazon Video Games                 & 50,907     & 17,389     & 452,004     \\
Anime                              & 60,970     & 11,197     & 6,250,866   \\
BookCrossing                       & 12,787     & 270,170    & 299,303     \\
Food                               & 22,018     & 226,590    & 830,889     \\
Goodreads                          & 260,025    & 2,021,053  & 14,651,363  \\
HotelRec                           & 2,029,381  & 365,013    & 21,660,081  \\
MovieLens                          & 162,541    & 59,047     & 24,753,332  \\
Netflix                            & 472,987    & 17,770     & 99,472,215  \\
Steam                              & 3,757      & 5,155      & 113,796     \\
WikiRec                            & 60,648     & 4,871,794  & 13,693,465  \\
Yelp (2021)                        & 287,113    & 150,346    & 4,350,452   \\ 
\hdashline
Amazon Beauty (2014) (*)           & 22,363     & 12,101     & 198,502     \\
Amazon Sports and Outdoors (2014) (*)  & 35,598     & 18,357     & 296,337     \\
Amazon Toys and Games (2014) (*)   & 19,412     & 11,924     & 167,597     \\
Yelp (2020) (*)                    & 30,431     & 20,033     & 316,354     \\ 
\hline
Total                              & 10,156,309 & 10,309,169 & 258,100,698 \\ \hline
\end{tabular}
\caption{Dataset statistics used for pre-training and fine-tuning. The asterisk (*) denotes datasets used exclusively in pre-training and final evaluation. For each of these four (*)-indicated datasets, we employ a train/validation/test split from previous works \cite{p5, genrec}, but we remove users and interactions from the training split if they appear in the validation/test split. This ensures that the validation/test data does not leak into the training data. Note that for these four datasets, we report our final evaluation scores on the test split, while the training split is only used for pre-training RecGPT-7B to mimic real-world scenarios. In other words, we do not use the training/validation split for supervised fine-tuning of instruction following.
Note that some datasets have two versions associated with different publication times (e.g., Amazon and Yelp). To maintain consistent test data with previous works, we retain the older versions (2014 for Amazon and 2020 for Yelp) for testing purposes and use the newer versions (2018 for Amazon and 2021 for Yelp) to enrich our pre-training data. We filter out overlapped users along with their interactions in the newer dataset to prevent duplication and data leakage.}
\label{tab:data-train}
\end{table*}

\section{Data format used in training and inference}
\label{sec:apd-prompt}

We present the prompt templates used in our work. 
Note that in both pre-training and fine-tuning phases, if a user has a long interaction history with many items (i.e., the number of tokens exceeds the max\_seq\_length of 2048), we pre-split the history into smaller chunks with a similar number of items, ensuring that the number of tokens in each chunk is smaller than 2048. Each chunk is then considered a separate user's interaction history. 

\subsection{Data format used in pre-training phase}

\textbf{Amazon}
\begin{spverbatim}
Given the interaction history of a user with products as follows:
Title: {title}; Brand: {brand}; Review: {review}; Rating: {rating}/5.0
...
Title: {title}; Brand: {brand}; Review: {review}; Rating: {rating}/5.0
\end{spverbatim}
\textbf{Amazon Books}
\begin{spverbatim}
Given the interaction history of a user with books as follows:
Title: {title}; Brand: {brand}; Review: {review}; Rating: {rating}/5.0
...
Title: {title}; Brand: {brand}; Review: {review}; Rating: {rating}/5.0
\end{spverbatim}

\medskip 

\textbf{Anime}
\begin{spverbatim}
Given the interaction history of a user with movies/shows as follows:
Title: {title}; Genres: {genres}; Rating: {rating}/10.0
...
Title: {title}; Genres: {genres}; Rating: {rating}/10.0
\end{spverbatim}
\textbf{BookCrossing}
\begin{spverbatim}
Given the interaction history of a user with books as follows:
Title: {title}; Author: {author}; Rating: {rating}/10.0
...
Title: {title}; Author: {author}; Rating: {rating}/10.0
\end{spverbatim}
\textbf{Food}
\begin{spverbatim}
Given the interaction history of a user with food recipes as follows:
Title: {title}; Review: {review_text}; Rating: {rating}/5.0
...
Title: {title}; Review: {review_text}; Rating: {rating}/5.0
\end{spverbatim}
\textbf{Goodreads}
\begin{spverbatim}
Given the interaction history of a user with books as follows:
Title: {title}; Author: {author}; Genres: {genres}; Review: {review_text}; Rating: {rating}/5.0
...
Title: {title}; Author: {author}; Genres: {genres}; Review: {review_text}; Rating: {rating}/5.0
\end{spverbatim}
\textbf{HotelRec}
\begin{spverbatim}
Given the interaction history of a user with hotels as follows:
Title: {title}; City: {city}; Review: {review_text}; Rating: {rating}/5.0
...
Title: {title}; City: {city}; Review: {review_text}; Rating: {rating}/5.0
\end{spverbatim}
\textbf{MovieLens}
\begin{spverbatim}
Given the interaction history of a user with movies/shows as follows:
Title: {title}; Genres: {genres}; Rating: {rating}/5.0
...
Title: {title}; Genres: {genres}; Rating: {rating}/5.0
\end{spverbatim}
\textbf{Netflix}
\begin{spverbatim}
Given the interaction history of a user with movies/shows as follows:
Title: {title}; Rating: {rating}/5.0
...
Title: {title}; Rating: {rating}/5.0
\end{spverbatim}
\textbf{Steam}
\begin{spverbatim}
Given the interaction history of a user with video games as follows:
Title: {title}
...
...Title: {title}

\end{spverbatim}
\textbf{WikiRec}
\begin{spverbatim}
Given the interaction history of a user with Wikipedia articles as follows:
Title: {title}; Description: {description}
...
Title: {title}; Description: {description}
\end{spverbatim}
\textbf{Yelp}
\begin{spverbatim}
Given the interaction history of a user with businesses as follows:
Title: {title}; City: {city}; Review: {review_text}; Rating: {rating}/5.0
...
Title: {title}; City: {city}; Review: {review_text}; Rating: {rating}/5.0
\end{spverbatim}

\subsection{Data format used in fine-tuning and inference}

\subsubsection{Rating prediction task}
\textbf{Amazon}
\begin{spverbatim}
### Instruction:
Predict rating for the last item. 
Given the interaction history of a user with products as follows:
Title: {title}; Brand: {brand}; Review: {review}; Rating: {rating}/5.0
...
Title: {title}; Brand: {brand}; Review: {review}; Rating:
### Response:
{rating}/5.0
\end{spverbatim}
\textbf{Amazon Books}
\begin{spverbatim}
### Instruction:
Predict rating for the last item.
Given the interaction history of a user with books as follows:
Title: {title}; Author: {author}; Review: {review}; Rating: {rating}/5.0
...
Title: {title}; Author: {author}; Review: {review}; Rating:
### Response:
{rating}/5.0
\end{spverbatim}
\textbf{Anime}
\begin{spverbatim}
### Instruction:
Predict rating for the last item.
Given the interaction history of a user with movies/shows as follows:
Title: {title}; Genres: {genres}; Rating: {rating}/10.0
...
Title: {title}; Genres: {genres}; Rating:
### Response:
{rating}/10.0
\end{spverbatim}
\textbf{BookCrossing}
\begin{spverbatim}
### Instruction:
Predict rating for the last item.
Given the interaction history of a user with books as follows:
Title: {title}; Author: {author}; Rating: {rating}/10.0
...
Title: {title}; Author: {author}; Rating:
### Response:
{rating}/10.0
\end{spverbatim}
\textbf{Food}
\begin{spverbatim}
### Instruction:
Predict rating for the last item.
Given the interaction history of a user with food recipes as follows:
Title: {title}; Review: {review_text}; Rating: {rating}/5.0
...
Title: {title}; Review: {review_text}; Rating:
### Response:
{rating}/5.0
\end{spverbatim}
\textbf{Goodreads}
\begin{spverbatim}
### Instruction:
Predict rating for the last item.
Given the interaction history of a user with books as follows:
Title: {title}; Author: {author}; Genres: {genres}; Review: {review_text}; Rating: {rating}/5.0
...
Title: {title}; Author: {author}; Genres: {genres}; Review: {review_text}; Rating:
### Response:
{rating}/5.0
\end{spverbatim}
\textbf{HotelRec}
\begin{spverbatim}
### Instruction:
Predict rating for the last item.
Given the interaction history of a user with hotels as follows:
Title: {title}; City: {city}; Review: {review_text}; Rating: {rating}/5.0
...
Title: {title}; City: {city}; Review: {review_text}; Rating:
### Response:
{rating}/5.0
\end{spverbatim}
\textbf{MovieLens}
\begin{spverbatim}
### Instruction:
Predict rating for the last item.
Given the interaction history of a user with movies/shows as follows:
Title: {title}; Genres: {genres}; Rating: {rating}/5.0
..
Title: {title}; Genres: {genres}; Rating:
### Response:
{rating}/5.0
\end{spverbatim}
\textbf{Netflix}
\begin{spverbatim}
### Instruction:
Predict rating for the last item.
Given the interaction history of a user with movies/shows as follows:
Title: {title}; Rating: {rating}/5.0
...
Title: {title}; Rating:
### Response:
{rating}/5.0
\end{spverbatim}
\textbf{Yelp}
\begin{spverbatim}
### Instruction:
Predict rating for the last item.
Given the interaction history of a user with businesses as follows:
Title: {title}; City: {city}; Review: {review_text}; Rating: {rating}/5.0
...
Title: {title}; City: {city}; Review: {review_text}; Rating:
### Response:
{rating}/5.0
\end{spverbatim}

\subsubsection{Sequential recommendation task}
\textbf{Amazon}
\begin{spverbatim}
### Instruction:
Predict the next item. 
Given the interaction history of a user with products as follows:
Title: {title}; Brand: {brand}
...
Title: {title}; Brand: {brand}
### Response:
Title: {title}; Brand: {brand}
\end{spverbatim}
\textbf{Amazon Books}
\begin{spverbatim}
### Instruction:
Predict the next item.
Given the interaction history of a user with books as follows:
Title: {title}; Author: {brand}; 
...
Title: {title}; Author: {brand};
### Response:
Title: {title}; Author: {brand};
\end{spverbatim}
\textbf{Anime}
\begin{spverbatim}
### Instruction:
Predict the next item.
Given the interaction history of a user with movies/shows as follows:
Title: {title}; Genres: {genres}
...
Title: {title}; Genres: {genres}
### Response:
Title: {title}; Genres: {genres}
\end{spverbatim}
\textbf{BookCrossing}
\begin{spverbatim}
### Instruction:
Predict the next item.
Given the interaction history of a user with books as follows:
Title: {title}; Author: {author}
...
Title: {title}; Author: {author}
### Response:
Title: {title}; Author: {author}
\end{spverbatim}
\textbf{Food}
\begin{spverbatim}
### Instruction:
Predict the next item.
Given the interaction history of a user with food recipes as follows:
Title: {title}
...
Title: {title}
### Response:
Title: {title}
\end{spverbatim}
\textbf{Goodreads}
\begin{spverbatim}
### Instruction:
Predict the next item.
Given the interaction history of a user with books as follows:
Title: {title}; Author: {author}; Genres: {genres}
...
Title: {title}; Author: {author}; Genres: {genres}
### Response:
Title: {title}; Author: {author}
\end{spverbatim}
\textbf{HotelRec}
\begin{spverbatim}
### Instruction:
Predict the next item.
Given the interaction history of a user with hotels as follows:
Title: {title}; City: {city}
...
Title: {title}; City: {city}
### Response:
Title: {title}; City: {city}
\end{spverbatim}
\textbf{MovieLens}
\begin{spverbatim}
### Instruction:
Predict the next item.
Given the interaction history of a user with movies/shows as follows:
Title: {title}; Genres: {genres}
..
Title: {title}; Genres: {genres}
### Response:
Title: {title}
\end{spverbatim}
\textbf{Netflix}
\begin{spverbatim}
### Instruction:
Predict the next item.
Given the interaction history of a user with movies/shows as follows:
Title: {title}
...
Title: {title}
### Response:
Title: {title}
\end{spverbatim}
\textbf{Steam}
\begin{spverbatim}
### Instruction:
Predict the next item.
Given the interaction history of a user with video games as follows:
Title: {title}
...
Title: {title}
### Response:
Title: {title}
\end{spverbatim}
\textbf{WikiRec}
\begin{spverbatim}
### Instruction:
Predict the next item.
Given the interaction history of a user with Wikipedia articles as follows:
Title: {title}; Description: {description}
...
Title: {title}; Description: {description}
### Response:
Title: {title}; Description: {description}
\end{spverbatim}
\textbf{Yelp}
\begin{spverbatim}
### Instruction:
Predict the next item.
Given the interaction history of a user with businesses as follows:
Title: {title}; City: {city}
...
Title: {title}; City: {city}
### Response:
Title: {title}; City: {city}
\end{spverbatim}

\end{document}